\documentclass[twocolumn,showpacs,prl]{revtex4}
\usepackage{amsmath}
\usepackage{graphicx}
\usepackage{dcolumn}
\usepackage{bm}

\begin{document}

\title{Transport properties of granular metals at low temperatures}
\author{I.~S.~Beloborodov}
\affiliation{Bell Laboratories, Lucent Technologies, Murray Hill, New Jersey 07974}
\affiliation{Department of Physics, University of Colorado, CB 390, Boulder, Colorado
80390}
\author{K.~B.~Efetov}
\affiliation{Theoretische Physik III, Ruhr-Universit\"{a}t Bochum, Germany}
\affiliation{L.~D.~Landau Institute for Theoretical Physics, 117940, Moscow, Russia}
\author{A.~V.~Lopatin}
\affiliation{Materials Science Division, Argonne National Laboratory, Argonne, Illinois
60439}
\author{V.~M.~Vinokur}
\affiliation{Materials Science Division, Argonne National Laboratory, Argonne, Illinois
60439}
\date{\today}
\pacs{73.23Hk, 73.22Lp, 71.30.+h}

\begin{abstract}
We investigate transport in a granular metallic system at large
tunneling conductance between the grains, $g_{T}\gg 1$. 
We show that at low temperatures, $%
T\leq g_{T}\delta $, where $\delta $ is the single mean energy level spacing
in a grain, the coherent electron motion at large distances dominates the
physics, contrary to the high temperature ($T>g_{T}\delta $) behavior where
conductivity is controlled by the scales of the order of the grain size.
The conductivity of one and two dimensional granular metals, in the low 
temperature regime, decays with decreasing temperature in the same 
manner as that in homogeneous disordered metals, indicating thus an 
insulating behavior. However, even in this temperature regime the 
granular structure remains
important and there is an additional contribution to conductivity coming
from short distances. Due to this contribution the metal-insulator
transition in three dimensions occurs at the value of tunnel conductance $%
g_{T}^{C}=(1/6\pi )\ln (E_{C}/\delta )$, where $E_{C}$ is the charging
energy of an isolated grain, and not at the generally expected $%
g_{T}^{C}\propto 1$. Corrections to the density of states of granular metals
due to the electron-electron interaction are calculated. Our results compare
favorably with the logarithmic dependence of resistivity in the high-$T_{c}$
cuprate superconductors indicating that these materials may have a granular
structure.
\end{abstract}

\maketitle

A great deal of research in the current mesoscopic physics focuses on
understanding properties of granular metals(see ~\cite{experiment,
Simon,Efetov02}). The interest is motivated by the fact that while their
properties are generic for a wealth of strongly correlated systems with
disorder, granular metals offer a unique experimentally accessible tunable
system where both the interaction strength and degree of disorder can be
controlled.

The key phenomenon revealing the most of the underlying physics is
transport, where the effects of interactions play a crucial role. The
processes of electron tunneling from grain to grain that govern electron
transfer, are accompanied by charging the grains involved after each
electron hop to another grain. This may lead to a Coulomb blockade, and one
justly expects this effect to be of the prime importance at least in the
limit of weak coupling. It makes it thus clear, on a qualitative level, that
it is the interplay between the the grain-to-grain coupling and the
electron-electron Coulomb interaction that controls transport properties of
granular metals; yet, despite the significant efforts expended, a
quantitative theory of transport in metallic granular systems is still
lacking.

A step towards formulation such a theory was made recently in [~%
\onlinecite{Efetov02}]. It was shown that depending on the dimensionless
tunneling conductance $g_{T}$ one observes either exponential-, at $g_{T}\ll
1$, or logarithmic, at $g_{T}\gg 1$ temperature dependence of conductivity.
The consideration in [~\onlinecite{Efetov02}] was based on the approach
developed by Ambegaokar, Eckern and Sch\"{o}n (AES)~\cite{AES} for tunnel
junctions. This technique however, as shown in ~\cite{Efetov}, applies only
at temperatures $T > g_T \delta$, where $\delta$ is the mean energy level 
spacing in a single grain, in this regime the electron coherence does not 
extend beyond the grain size. 
At low temperature region, $T\leq g_{T}\delta $, the effects of the
electron coherent motion at distances much exceeding the single grain size $a
$ must be included, thus this important regime is not described by the AES
approach ~\cite{Efetov}.

Although experimentally the low temperature regime is well within the
experimental reach ~\cite{experiment,Simon}, it has never been addressed
theoretically so far. The important question whether the system is a metal
or becomes an insulator, in other words, whether the conductivity of the
granular metals at large conductances remain finite in the limit of $%
T\rightarrow 0$ is still open.

In this Letter we investigate the low-temperature conductivity of granular
samples focusing on the case of large tunneling conductance between the
grains, $g_{T}\gg 1$. To this end we develop a technique that goes beyond
the AES approach and includes effects of coherent electron motion at
distances larger than the size of the grain. Without the Coulomb interaction
the granular system would be a good metal in the limit, $g_{T}\gg 1$, and
our task is to include the charging effects into the theory. We find that at
temperatures, $T\leq g_{T}\delta $ properties of the granular metal depend
on the dimensionality of the array, and corrections to the conductivity and
density of states due to the effects of Coulomb interaction are similar to
those obtained in Ref.~\onlinecite{Altshuler} for a homogeneous metal. Thus
at low temperatures the systems behaves essentially as a homogeneous metal
contrasting the case of large temperatures, $T\gg g_{T}\delta $ considered
in Ref.~\cite{Efetov02}.

This means that at large conductances the 3$D$ system is a good metal. On
the other hand, at $g_T \ll 1$ a granular sample is in the insulating state.
Therefore a 3D system should exhibit a metal-insulator transition at the
critical value of the conductance $g_{T},$ such that samples with
conductances $g_{T}>g_{T}^{C}$ are metals and their conductivity remains
finite at $T\rightarrow 0$ while samples with $g_{T}<g_{T}^{C}$ are
insulators and their conductivity vanishes at $T\rightarrow 0.$

The main results of our work are as follows: (i) We estimate the critical
value $g_{T}^{C}$ of the tunnel conductance at which the metal-insulator 
transition in 3$D$ occurs as 
\begin{equation}  \label{gC}
g_{T}^{C}=(1/6\pi )\ln(E_{C}/\delta ),
\end{equation}
where $E_{C}$ is the charging energy of an isolated grain. (ii) We find the
expression for the conductivity of a granular metal that includes
corrections due to Coulomb interaction and holds for all temperatures as
long as these corrections are small. The corresponding answer can be
conveniently written separating the correction due to the contribution from
the large energy scales $\varepsilon> g_T\delta $ from that coming from the
low energy scales $\varepsilon< g_T\delta.$ Denoting corrections as $\delta
\sigma_{1}$ and $\delta \sigma _{2}$ respectively we have 
\begin{subequations}
\label{result0}
\begin{equation}
\sigma =\sigma _{0}+\delta \sigma _{1}+\delta \sigma _{2},
\label{mainresult1}
\end{equation}
where $\sigma _{0}=2 e^{2}g_{T}a^{2-d}$, with $a$ being the size of the
single grain is the classical Drude conductivity for a granular metal (spin
included). Correction $\delta \sigma_{1}$ in Eq.~(\ref{mainresult1})
contains the dimensionality of the array $d$ only as a coefficient and is
given by the following expression~\cite{Efetov02}, 
\begin{equation}
\frac{\delta \sigma _{1}}{\sigma _{0}}=-{\frac{{1}}{{\ 2\pi dg_{T}}}}\,\ln %
\left[ {\frac{{g_{T}E_{C}}}{\max {(T,g_{T}\delta )}}}\right] .
\label{mainresult3}
\end{equation}
On the contrary the correction $\delta \sigma _{2}$ in Eq.~(\ref{mainresult1}%
) that is important only at temperatures $T<\delta g_T$ strongly depends on
the dimensionality of the array 
\begin{equation}  \label{mainresult4}
\frac{\delta \sigma _{2}}{\sigma _{0}}=\left\{ 
\begin{array}{lr}
{\frac{{\alpha }}{{12\pi ^{2}g_{T}}}}\sqrt{{\frac{{T}}{{g_{T}\delta }}}} 
\hspace{1.6cm} D=3, &  \\ 
-\frac{1}{4\pi ^{2}g_{T}}\ln \frac{g_{T}\delta }{T}\hspace{1.4cm} D=2, &  \\ 
-{\frac{{\beta }}{{4\pi }}}\sqrt{{\frac{{\ \delta }}{{Tg_{T}}}}} \hspace{
1.9cm} D=1. & 
\end{array}
\right.
\end{equation}
Here $\alpha =\int_{0}^{\infty }dx\,x^{-1/2}[1-\coth (x)+x/\sinh
^{2}(x)]\approx 1.83$ and $\beta =\int_{0}^{\infty }dx\,x^{-3/2}\,[\coth
(x)-x/\sinh ^{2}(x)]\approx 3.13$ are the numerical constants. For a $3D$
granular system a temperature independent term of the order $1/g_{T}$ has
been subtracted in the first line in Eq.~(\ref{mainresult4}).

Corrections $\delta \sigma _{1}$ and $\delta \sigma _{2}$ are of a different
origin: the correction $\delta \sigma _{1}$ comes from the large energy
scales, $\varepsilon>g_{T}\delta $ where the granular structure of the array
dominates the physics. The fact that this correction is essentially
independent of the dimentionality $d$ means that the tunneling of electrons
with energies $\varepsilon> g_{T}\delta $ can be considered as incoherent.
On the other hand, correction $\delta \sigma _{2}$ in Eq.~(\ref{mainresult4}%
) is similar to that obtained for homogeneous metals long ago~\cite%
{Altshuler} and comes from the low energy scales, $\varepsilon \leq
g_{T}\delta $, where the coherent electron motion on the scales larger than
the grain size $a$ dominates the physics.

It is important to note that in the low temperature regime all temperate
dependence of conductivity comes from the correction $\delta\sigma_2.$ At
the same time, in this regime the correction $\delta \sigma_1,$ though being
temperature independent, still exists and can be even larger than $%
\delta\sigma_2.$

When deriving Eqs.~(\ref{result0}) we neglected possible weak localization
corrections that may originate from quantum interference of electron waves.
This approximation is legitimate if a magnetic field is applied as in Ref.~%
\cite{experiment} or dephasing is strong due to inelastic processes.

Now we turn to the description of our model and  the derivation of Eqs.~(\ref%
{result0}):  We consider a $d-$dimensional array of metallic grains with the
Coulomb interaction between electrons. The motion of electrons inside the
grains is diffusive and they can tunnel from grain to grain. In principle,
the grains can be clean such that electrons scatter mainly on grain
surfaces. We assume that the sample in the absence of the Coulomb
interaction would be a good metal. For large tunneling conductance we may
also neglect the nonperturbative charging effects (discretness of the 
electron charge)~\cite{Beloborodov03},  which give an exponentially small
(as $\exp(-\# g_{T})$) contribution to the conductivity. Although we assume
that the dimensionless tunneling conductance $g_{T}$ is large, it should be
still smaller than the grain conductance, $g_{0},$ such that $g_{T}<g_{0}.$
This inequality means that the granular structure is still important and the
main contribution to the macroscopic resistivity comes from the contacts
between the grains. The limit of the homogeneous metal is reached when the
grain conductance, $g_{0}$ becomes of the order of tunneling conductance, $%
g_{0}\sim g_{T}$.

The system of weakly coupled metallic grains can be described by the
Hamiltonian 
\end{subequations}
\begin{equation}
\hat{H}=\hat{H}_{0}+\hat{H}_{c}+\sum_{ij}\,t_{ij}\,[\,\hat{\psi}^{\dagger
}(r_{i})\,\hat{\psi}(r_{j})+\hat{\psi}^{\dagger }(r_{j})\,\hat{\psi}%
(r_{i})\,],  \label{hamiltonian}
\end{equation}
where $t_{ij}$ is the tunneling matrix element corresponding to the points
of contact $r_{i}$ and $r_{j}$ of $i$-th and $j-$th grains. The Hamiltonian $
\hat{H}_{0}$ in Eq.~(\ref{hamiltonian}) describes noninteracting isolated
disordered grains. The term $\hat{H}_{c}$ describes the Coulomb interaction
inside and between the grains. It has the following form 
\begin{equation}
\hat{H}_{c}={\frac{{\ e^{2}}}{{\ 2}}}\,\sum_{ij}\,\hat{n}_{i}\,C_{ij}^{-1}\,%
\hat{n}_{j},  \label{Coulomb}
\end{equation}
where $C_{ij}$ is the capacitance matrix and $\hat{n}_{i}$ is the operator
of electrons number in the $i$-th grain. In the regime under consideration
one can neglect the coordinate dependence of a single grain diffusion
propagator. The electron hopping between the grains can be included using
the diagrammatic technique developed 
in Refs.~\onlinecite{Beloborodov99,Efetov}.

The conductivity of the granular metals is given by the analytical
continuation of the Matsubara current-current correlator. In the absence of
the electron-electron interaction the conductivity is represented by the
diagram (a) in Fig.~1 that results in high temperature (Drude) conductivity $%
\sigma _{0}$ which is defined below Eq.~(\ref{mainresult1}). First order
interaction corrections to the conductivity are given by the diagrams (b-e)
in Fig.~1. These diagrams are analogous to ones considered in Ref.~%
\onlinecite{Altshuler} for the correction to the conductivity of homogeneous
metals. We consider the contributions from diagrams (b,c) and (d,e)
separately: The sum of the diagrams (b,c) results in the following
correction to the conductivity 
\begin{equation}
\frac{\delta \sigma _{1}}{\sigma _{0}}=-{\frac{1}{{2\pi dg_{T}}}} {\rm Im} 
\sum_{\mathbf{q}}\int d\omega \,\gamma (\omega )\,\varepsilon _{\mathbf{q}}\,%
\, \tilde{V}(\omega ,\mathbf{q}).  \label{Diagrams12}
\end{equation}%
where $\gamma (\omega )={\frac{{d}}{{d\omega }}} \omega \coth {\frac{{\omega 
}}{{2T}}},$ $\varepsilon _{q}=2 g_{T}\sum_{\mathbf{a}}(1-\cos \mathbf{qa})$
with $\mathbf{a}$ being the lattice vectors and 
\begin{equation}
\tilde{V}(\omega ,\mathbf{q})={\frac{{\ 2\,E_{C}(\mathbf{q})}}{{(\varepsilon
_{\mathbf{q}}\delta -i\omega )\,(4\,\varepsilon _{\mathbf{q}}E_{C}(\mathbf{q}%
)-i\omega )}}}.  \label{effectivinteraction}
\end{equation}
Here the charging energy $E_{C}(\mathbf{q})=e^{2}/2C(\mathbf{q})$ is
expressed in terms to the Fourier transform of the capacitance matrix $C(%
\mathbf{q})$. Performing the integration over the frequency and summing over
the quasimomentum $\mathbf{q}$ in Eq.~(\ref{Diagrams12}) with the
logarithmic accuracy we obtain the correction~(\ref{mainresult3}). One can
see from Eq.~(\ref{Diagrams12}) that the contribution $\delta \sigma _{1}$
in Eq.~(\ref{mainresult3}) comes from the large energy scales, $\varepsilon
>g_{T}\delta $ such that at low temperatures the logarithm is cut off on the
energy scale $g_{T}\delta .$ 
\begin{figure}[tbp]
\resizebox{.43\textwidth}{!}{\includegraphics{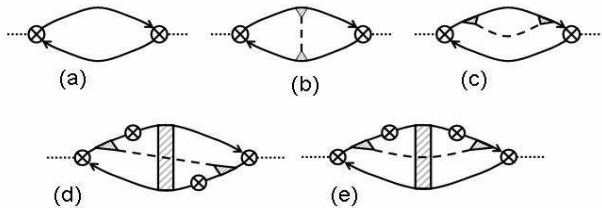}} \vspace{0.6cm}
\caption{Diagrams describing the conductivity of granular metals: the
diagram (a) corresponds to $\protect\sigma _{0}$ in Eq.~(\ref{mainresult1})
and it is the analog of Drude conductivity. Diagrams (b)-(e) describing
first order correction to the conductivity of granular metals due to
electron-electron interaction. The solid lines denote the propagator of
electrons and the dashed lines describe effective screened electron-electron
propagator. The tunneling vertices are described by the circles. The sum of
the diagrams (b) and (c) results in the conductivity correction $\protect%
\delta \protect\sigma _{1}$ in Eq.~(\ref{mainresult1}). The other two
diagrams, (d) and (e) result in the correction $\protect\delta \protect%
\sigma _{2}$. }
\label{fig:2}
\end{figure}

To obtain the total correction to the conductivity of granular metal the two
other diagrams, (d) and (e) in Fig.~1 should be taken into account. These
diagrams result in the following contribution to the conductivity 
\begin{equation}
{\frac{{\delta \sigma _{2}}}{{\ \sigma _{0}}}}=-{\frac{{\ 2g_{T}\delta }}{{\
\pi d}}}\sum_{\mathbf{q}}\int d\omega \,\gamma (\omega )\,{\text{Im}\frac{{%
\tilde{V}(\omega ,\mathbf{{q})\sum_{a}\sin ^{2}({q}a)}}}{{\varepsilon _{%
\mathbf{q}}\delta -i\omega }}}.  \label{sigma2}
\end{equation}%
In contrast to the contribution $\delta \sigma _{1}$ in Eq.~(\ref{Diagrams12}%
), the main contribution to the sum over the quasimomentum $\mathbf{q}$ in
Eq.~(\ref{sigma2}) comes from the low momenta, $q\ll 1/a$. In this regime
the capacitance matrix, $C(\mathbf{q)}$ in Eqs.~(\ref{effectivinteraction})
and (\ref{sigma2}) has the following asymptotic form 
\begin{equation}
C^{-1}(\mathbf{q})=\frac{2}{a^{d}}\,\left\{ 
\begin{array}{rl}
& \ln (1/qa)\hspace{0.7cm} D=1, \\ 
& \pi /q\hspace{1.4cm} D=2, \\ 
& 2\pi /q^{2}\hspace{1.1cm} D=3.%
\end{array}%
\right.  \label{capacitance}
\end{equation}%
Substituting Eq.~(\ref{capacitance}) into Eq.~(\ref{effectivinteraction}),
integrating over the frequency and summing over the quasimomentum $\mathbf{q}
$ in Eq.~(\ref{sigma2}) we obtain the result for the correction $\delta
\sigma _{2}$ in Eq.~(\ref{mainresult4}). This type of corrections to the
conductivity cannot be obtained from the effective AES action.

Comparing our results in Eqs.~(\ref{result0}) with those obtained in Ref.~%
\onlinecite{Efetov02} using the AES functional we see that the correction to
the conductivity obtained in Ref.~\onlinecite{Efetov02} is equivalent to the
correction $\delta \sigma _{1}$ in Eq.~(\ref{mainresult1}), which
corresponds in our approach to the sum of diagrams (b) and (c) in Fig.~1.
The correction $\delta \sigma _{2}$ in Eq.~(\ref{mainresult1}) becomes
important only at low temperatures, $T<g_{T}\delta $ where AES functional is
not applicable. While in our approach both corrections to the conductivity
must be small $\delta \sigma _1, \delta \sigma _2 \ll \sigma _{0}$ the
method of Ref.~\onlinecite{Efetov02} gives a possibility to show that for $%
T\gg g_{T}\delta $ the dependence of the conductivity is logarithmic so long
as $\sigma /e^{2} a^{2-d} \gg 1$.

It follows from Eq.~(\ref{mainresult4}) that at low temperatures, $%
T<g_{T}\delta ,$ for a $3D$ granular array, there are no essential
corrections to the conductivity coming from the low energies since the
correction $\delta \sigma _{2}$ is always small. This means that the result
for the renormalized conductance, $\tilde{g}_{T}$ of Ref.~%
\onlinecite{Efetov02} for $3D$ samples within the logarithmic accuracy can
be written in the following form 
\begin{equation}
\tilde{g}_{T}(T)=g_{T}-{\frac{{1}}{{6\pi }}}\ln \left[ {\frac{{g_{T}E_{C}}}{%
\max {(\tilde{g}_{T}\delta ,T)}}}\right] ,  \label{RGgen}
\end{equation}
such that it is valid for \textit{all} temperatures as long as the
renormalized conductance, $\tilde{g}_{T}\gg 1$. One can see from Eq.~(\ref%
{RGgen}) that for bare conductance, $g_{T}\gg (1/6\pi )\ln
(g_{T}E_{C}/\delta )$ the renormalized conductance, $\tilde{g}_{T}$ is
always large and the system remains metallic down to zero temperatures. In
the opposite limit $g_{T}<(1/6\pi )\ln (g_{T}E_{C}/\delta )$, the system
flows when decreasing the temperature to the strong coupling regime, $\tilde{%
g}_{T}\sim 1$ that indicates the onset of the insulating phase. We see that
with the logarithmic accuracy the critical value of the conductance $%
g_{T}^{C}$ is given by Eq.~(\ref{gC}).

In a similar way we can obtain interaction corrections to the density of
states (DOS) of granular metal 
\begin{equation}
\frac{\delta \nu (\varepsilon )}{\nu _{0}}=-{\frac{{1}}{{4\pi }}}\sum_{%
\mathbf{q}}\text{Im}\,\int d\omega \,{\frac{{\tanh [(\varepsilon -\omega
)/2T]}}{{(\varepsilon _{\mathbf{q}}\delta -i\omega )[\varepsilon _{\mathbf{q}
}-i\omega /4E_{C}(\mathbf{q})]}}}.  \label{density_of_states}
\end{equation}
Here $\nu _{0}$ is the DOS for noninteracting electrons, $\varepsilon _{%
\mathbf{q}}$ and $E_{C}(\mathbf{q})$ were defined below Eqs.~(\ref%
{Diagrams12} ) and (\ref{effectivinteraction}) respectively. Using Eq.~(\ref%
{density_of_states}) for a $3D$ granular array we obtain 
\begin{subequations}
\label{DOS}
\begin{equation}
{\frac{{\delta \nu _{3}}}{{\nu _{0}}}}=-{\frac{{A}}{{2\pi g_{T}}}}\ln \left[ 
{\frac{{E_{C}g_{T}}}{\max {(T,g_{T}\delta )}}}\right],  \label{density_3D}
\end{equation}
where $A=g_{T} a^3 \int d^{3}q\,/(2\pi )^{3}\,\varepsilon _{\mathbf{q}}^{-1}$.
For temperatures, $ T\gg g_{T}\delta $ the correction to the DOS (\ref%
{density_3D}) coincides with the one obtained in Ref.~\cite{Efetov02} using
AES approach. It follows from Eq.~(\ref{density_3D}) that for a $3D$ array
of grains, as in case with conductivity, the main contribution to the DOS
comes from the large energy scales, $\varepsilon > g_{T}\delta$.

Using Eq.~(\ref{density_of_states}) for a $2D$ array we obtain the following
result for the correction to the DOS 
\begin{equation}  \label{Density_d=3}
\frac{\delta \nu _{2}}{\nu _{0}}=-\frac{1}{16g_{T}\pi ^{2}}\left\{ 
\begin{array}{lr}
2\ln ^{2}{\frac{{g_{T}E_{C}}}{{T}}}\hspace{2.2cm}T\gg g_{T}\delta , &  \\ 
\ln {\frac{{g_{T}\delta }}{{T}}}\ln {\frac{{gE_{0}^{4}}}{{T\delta ^{3}}}}
+2\ln ^{2}{\frac{{E_{0}}}{{\delta }}}\hspace{0.3cm}T\ll g_{T}\delta . & 
\end{array}
\right.
\end{equation}
Using the relation between the tunneling conductance 
and the diffusion coefficient, $D=g_{T}a^{2}\delta $ one can see that the
temperature dependence of the DOS for $T\ll g_{T}\delta $ given by Eq.~(\ref%
{Density_d=3}) coincides up to the temperature independent constant with the
result for the correction to the DOS of the homogeneous metal~\cite%
{Altshuler}.

A logarithmic dependence of resistivity on temperature was recently found in
high-$T_{c}$ compounds $La_{2-y}Sr_{y}CuO_{4}$ and $%
Bi_{2}Sr_{2-x}La_{x}CuO_{6+\delta }$ in a very strong magnetic field \cite%
{boebinger,boebinger1}. A possible granularity of these samples was
suggested in Ref.~\cite{Efetov02}. Recently the microscopic granularity was
directly experimentally observed in the superconducting state of $%
Bi_{2}Sr_{2}CaCu_{2}0_{8+\delta }$ by the STM probe~\cite{Lang}. If we
accept that samples studied in \cite{boebinger,boebinger1} are indeed
microscopically granular, we can compare the results of the experiments with
our predictions. When doing so it is convenient to scale three dimensional
conductivity to the conductivity of CuO planes, $\sigma _{plane}.$ According
to our predictions 
\end{subequations}
\begin{equation}
{\frac{{d\sigma _{plane}}}{{d\ln T}}}={\frac{{e^{2}}}{{\pi \hbar }}}\,k , 
\label{experiment}
\end{equation}
where the coefficient $k=1/2\pi $ in the low temperature- and $k=1/d$ in the
high temperature regimes. While in the low temperature regime the
application of Eq.~(\ref{experiment}) is legitimate only under the
assumption that electrons in different CuO plane are incoherent, in the high
temperature regime the behavior of conductivity according to 
Eq.~(\ref{mainresult3}) is logarithmic for any dimension. In this regime the real
dimensionality $d$ should be replaced by $d=Z/2$, where $Z$ is the (average)
number of the contacts of each grain with all the adjacent grains.
Describing the data shown in Fig.~3 of Ref.~\cite{boebinger1} by our log
dependencies at temperatue $T\approx 5K$ we extract $k\simeq 0.4,$
for $Sr$ concentration of $y=0.08$ for $La_{2-y}Sr_{y}CuO_{4}$~\cite{curve}; for the $%
Bi_{2}Sr_{2-x}La_{x}CuO_{6+\delta }$ compound we find $k\simeq 0.2$ for $%
x=0.84$ $La$ concentration, and $k\simeq 0.3$ for $x=0.76.$ For each particular curve
the values $k$ extracted from Fig.~3 of Ref.~\cite{boebinger1}
increase with temperature (especially in case of $LSCO$), this is  in
a complete agreement with our results provided that the
``coherent-incoherent'' crossover occurs at about $T\sim 5K$. At higher
temperatures $k$ noticeably exceeds $1/2\pi $, supporting the idea of a
granularity of doped cuprates. The questions whether the high values of $
T_{c}$ in the superconductors are related to the granular structure and what
is the origin of this granularity are beyond the scope of the present work.

In conclusion, we have investigated transport properties of granular metals
at large tunneling conductance and obtained corrections to the conductivity,
Eqs.~(\ref{result0}) and DOS, Eqs.~(\ref{DOS}) due to electron-electron
interaction. We have shown that at temperatures, $T>g_{T}\delta $ the
granular structure of the array dominates the physics. On the contrary at
temperatures, $T\leq g_{T}\delta $ the large-scale coherent electron motion
is crucial. Comparison our results with experimental data supports the
assumption about a granular structure of doped high-$T_{c}$ cuprates.

We thank A.~Andreev, A.~Koshelev, A.~Larkin and K.~Matveev for useful
discussion of the results obtained. I.~B. thanks Materials Science Division
of Argonne National Laboratory for hospitality. K.~E. thanks German-Israeli
programs DIP and GIF for a support. This work was supported by the U.S.
Department of Energy, Office of Science through contract No. W-31-109-ENG-38
and by the A.P. Sloan and the Packard Foundations.


\vspace{-0.3cm}

\begin{thebibliography}{99}

\bibitem{experiment} A.~Gerber \textit{et al, } Phys. Rev. Lett.~\textbf{\ 78%
}, 4277 (1997).

\bibitem{Simon} R.~W.~Simon \textit{et al, } Phys. Rev. B \textbf{36}, 1962
(1987).

\bibitem{Efetov02} K.~B.~Efetov and A.~Tschersich, Europhys. Lett.~\textbf{\
59}, 114, (2002); cond-mat/0302257 (2003).

\bibitem{Efetov} I.~S.~Beloborodov \textit{et al, } Phys. Rev. B~\textbf{\ 63%
}, 115109 (2001).

\bibitem{AES} V.~Ambegaokar, U.~Eckern, and G.~Sch\"{o}n, Phys. Rev. Lett. 
\textbf{\ 48}, 1745 (1982).

\bibitem{Beloborodov03} I.~S.~Beloborodov and A.~V.~Andreev, in preparation.

\bibitem{Altshuler} B.~L.~Altshuler and A.~G.~Aronov, in \textit{\
Electron-Electron Interaction in Disordered Systems}, ed. by A.~L.~Efros and
M.~Pollak, North-Holland, Amsterdam (1985).

\bibitem{Beloborodov99} I.~S.~Beloborodov and K.~B.~Efetov, Phys. Rev. Lett.~%
\textbf{82}, 3332, (1999).

\bibitem{boebinger} Y.~Ando \textit{et al, } Phys. Rev. Lett.~\textbf{75},
4662, (1995); G.~S.~Boebinger \textit{et al}, Phys. Rev. Lett.~\textbf{77},
5417 (1996).

\bibitem{boebinger1} S.~Ono \textit{et al, } Phys. Rev. Lett.~\textbf{85},
638 (2000).

\bibitem{Lang} K.M. Lang \textit{et al, } Nature.~\textbf{415}, 412 (2002).

\bibitem{curve} This curve for $Bi_2Sr_{2-x}La_xCuO_{6+\delta}$, $x=0.15$
shows a very wide crossover at about 20K extending to low temperatures. Thus
there are just a very few points in the narrow temperature interval 1-3K
where our approach is applicable. This invalidates the comparison the data
with our results for this particular curve.
\end{thebibliography}
\end{document}